\def\year{2022}\relax
\documentclass[letterpaper]{article} 
\usepackage{aaai22}  
\usepackage{times}  
\usepackage{helvet}  
\usepackage{courier}  
\usepackage[hyphens]{url}  
\usepackage{graphicx} 
\usepackage{natbib}  
\usepackage{caption} 
\DeclareCaptionStyle{ruled}{labelfont=normalfont,labelsep=colon,strut=off} 
\usepackage{algorithm}
\usepackage{enumitem}
\usepackage{algorithmic}
\usepackage{multibib}
\usepackage{graphicx}
\usepackage{comment}
\usepackage{color,soul}

\usepackage{booktabs}
\usepackage{multirow}
\usepackage{url}
\usepackage{tabularx}

\usepackage{amsmath}
\usepackage{amsfonts} 
\usepackage{comment}
\usepackage{lipsum} 
\usepackage{amssymb}
\usepackage{xcolor} 
\usepackage{hyperref}
\usepackage{xstring}
\usepackage{color}   
\usepackage{wrapfig}

\usepackage{xcolor}

\usepackage{tcolorbox}

\usepackage{natbib}
\usepackage{newfloat}
\usepackage{listings}
\floatstyle{ruled}
\newfloat{listing}{tb}{lst}{}
\floatname{listing}{Listing}
\title{Bike Frames: Understanding the Implicit Portrayal of Cyclists in the News}
\author{
    Xingmeng Zhao$^1$, Dan Schumacher$^1$, Sashank Nalluri$^1$, Suahana Shrestha$^1$,\\Xavier Walton$^2$, and Anthony Rios$^1$
}
\affiliations{
    \textsuperscript{\rm 1}The University of Texas at San Antonio\\
    \textsuperscript{\rm 2}Alamo Colleges\\

}

\usepackage{bibentry}

\begin{document}
\maketitle

\begin{abstract}
Increasing cycling for transportation or recreation can boost health and reduce the environmental impacts of vehicles. However, news agencies' ideologies and reporting styles often influence public perception of cycling. For example, if news agencies overly report cycling accidents, it may make people perceive cyclists as ``dangerous,'' reducing the number of opting to cycle. Additionally, a decline in cycling can result in less government funding for safe infrastructure. In this paper, we develop a method for detecting the perceived perception of cyclists within news headlines. We introduce a new dataset called ``Bike Frames'' to accomplish this. The dataset consists of 31,480 news headlines and 1,500 annotations. Our focus is on analyzing 11,385 headlines from the United States. We also introduce the BikeFrame Chain-of-Code (CoC) framework to predict cyclist perception, identify accident-related headlines, and determine fault. This framework uses pseudocode for precise logic and integrates news agency bias analysis for improved predictions over traditional chain-of-thought reasoning in large language models. Our method substantially outperforms other methods, and most importantly, we find that incorporating news bias information substantially impacts performance, improving the average F1 from .739 to .815. Finally, we perform a comprehensive case study on US-based news headlines, finding reporting differences between news agencies and cycling-specific websites as well as differences in reporting depending on the gender of cyclists.
\textbf{WARNING: This paper contains descriptions of accidents and death.}\footnote{The code and data will be released on GitHub on acceptance.}
\end{abstract}

\section{INTRODUCTION}
Bicycling is a means of transportation and exercise that can reduce greenhouse gas emissions and improve public health. For instance, increasing pedestrian and bicycling trips, with a corresponding average decrease in automobile trip lengths by as little as one to three miles, can significantly affect emissions and fuel consumption~\cite{gotschi2008active}. Moreover, bicycling could reduce CO2 by six to fourteen million tons and reduce fuel consumption by 700 million to 1.6 billion gallons~\cite{gotschi2008active}. Bicycling can also provide health benefits such as reducing cardiovascular risk factors, the likelihood of coronary heart disease, general morbidity, mortality risk, cancer risk, and obesity~\cite{oja2011health}. Unfortunately, the public's perception of cycling---which can be influenced by how the news and social media portray cyclists---might impact both the number of cyclists and investment in cycling infrastructure by the community~\cite{berke2019bike}. \textcolor{black}{Individual reactions to the news, illustrating a broader interplay of personal values and actions, often guide their decisions and behaviors~\cite{chen2014understanding}. Hence, in this paper, we develop a method to analyze people's perception of cyclists and the language used in portraying cyclists in news headlines by introducing a new dataset ``Bike Frames'' and applying novel methods trained on this dataset to conduct a case study. This case study delves into extracting explicit and implicit information from headlines about how bicyclists---and motorcyclists as a comparison group---are portrayed.} 

\textcolor{black}{\citet{macmillan2016trends} introduce a framework for understanding the media's impact on public perceptions of cycling. Their analysis showed that while cycling trips in London doubled from 1992 to 2012, media coverage of cyclist fatalities increased 13-fold, a trend not observed in motorcyclist-related media coverage during the same period. This disproportionate coverage might create complex feedback loops affecting cycling growth, with the impact likely varying between cities. Therefore, analyzing public perception through news sources, like local agencies, is essential for understanding and addressing how bicyclists are portrayed, which is significant for the broader transportation and urban science research community. For instance, \citet{macmillan2017understanding} found that increasing cycling in cities is good for health and addressing climate change. Likewise, \citet{aldred2019barriers} found that media and public opposition were not reported as major issues for communities not considering new bicycling infrastructure, but they can substantially affect the release of funding for new cities that begin considering new infrastructure. However, ~\citet{aasvik2021cyclists} point out that prior work has mainly relied on surveys and qualitative analysis, making it a challenge to measure perceptions at scale, especially when comparing diverse communities.}

Previous research indicates that gender can influence factors like safety perception and home responsibilities, affecting cycling behaviors. For example, female cyclists~\footnote{Our discussion on gender, particularly the terms ``male'' and ``female,'' is rooted in the appearance of pronouns in cycling news headlines, not self-disclosed gender.} may be more concerned about road safety and more likely to be influenced by news broadcasts than males, especially if they perceive a heightened risk from such coverage~\cite{emond2009explaining}. Additionally, \citet{harris2006gender} discovered that women tend to be more risk-averse and feel more negative consequences of sharing the road with automobiles or other vehicles than men. These findings suggest that female cyclists may be more likely to favor off-road bike lanes separated from traffic~\cite{garrard2012women}. Understanding the interaction of gender on perception can help planners and policymakers develop strategies to promote cycling, particularly among women.

\begin{figure}[t]
    \centering
    \includegraphics[width=.55\linewidth]{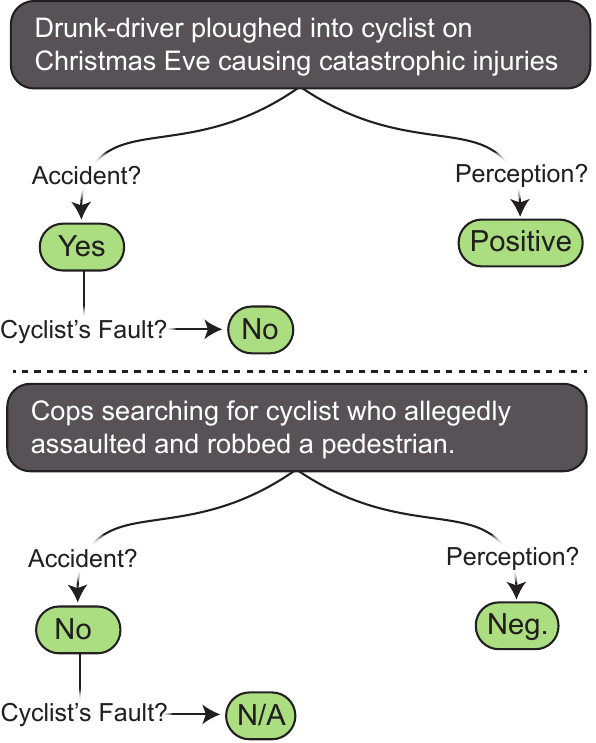}
    \caption{Understanding and explaining implicit views of cyclists requires reasoning. Hence, our ``Bike Frames'' aim to (1.) detect accident-related headlines, (2.) identify implicit suggestions of ``who was at fault'', and (3.) measure perception towards cyclists.\vspace{-1em}}
    \label{fig:overview}
\end{figure}


Overall, to better understand the perception of cyclists in news headlines, we introduce a novel Bike Frames Corpus, methodology, and analysis approaches that can extract the perception portrayed by news headlines using machine learning algorithms. We summarize the task our dataset addresses in Figure~\ref{fig:overview}. Specifically, we measure two aspects: the perceived view of the writer's feelings about the bicyclist and information about whether or not the bicyclist was in an accident. In the example ``\textit{Drunk-driver plowed into the cyclist},'' we can understand that the headline is related to a driver-bicyclist accident and that the driver was probably at fault. We may perceive that the writer expresses sympathy towards the cyclist (i.e., sees the cyclist in a positive light) given their explicit remark about the driver's initial condition (intoxicated) and through the strong verb ``\textit{plowed}.'' In contrast, if the headline said ``\textit{Cyclist allegedly hit by a driver},'' the text would still discuss an accident. However, we may perceive the writer as having a more positive perception of the driver, reducing the positive perception of the bicyclist. Therefore, we developed the dataset to measure the public perceptions of bicyclists and motorcyclists from news headlines for direct comparison. Specifically, the approach uncovers readers’ sentiment patterns toward cycling-related news articles. 

We explore using large language models (LLMs) to predict perception and fault in news headlines. Substantial research uses in-context examples and chain-of-thought reasoning to improve prediction~\cite{wei2022chain, chen2023program}. However, Traditional methods like Chain-of-Thought~\cite{wei2022chain} and Program-of-Thought~\cite{chen2023program} process logic in a single operation, limiting expressiveness and efficiency. Inspired by \citet{li2023chain} and \citet{chae2024language}, we introduce the BikeFrame Chain-of-Code (CoC) framework to enhance the reasoning of large language models (LLMs) by using pseudocode for precise logic management.  Our approach breaks down complex problems into intermediate reasoning steps, integrating news agency bias analysis to predict accident occurrences, identify fault, and assess cyclists' perceptions. This method uses pseudocode prompts to guide LLMs, with intermediate \texttt{print()} statements to clarify decision rationales. The use of news agency bias is highly informative. Research has shown that news agencies produce biased headlines~\cite{weatherly2007perceptions}. Hence, we explore whether asking LLMs to reason about these biases will improve the predictive performance of classifying cyclist perception.

\textcolor{black}{In summary, our contribution is three-fold. First, we introduce a new dataset for understanding the perception of cyclists in news headlines that may influence readers' perceptions of bicyclists. Second, we developed a novel prompting method called BikeFrame Chain-of-Code, which incorporates pseudocode logic with news agency bias reasoning tasks, substantially improving over traditional prompting methods. Third, we perform a comprehensive case study on important aspects of transportation---the portrayal of bicyclists in media---which has not been the main focus of prior research. Our work expands on prior media framing biases~\cite{levin1998all,rho2018fostering}. We investigate gender biases in reporting cycling accidents, comparing portrayals of male versus female cyclists.} 

\section{RELATED WORK}
In this section, we describe four main areas of research related to this paper: Bicycle and general travel infrastructure research, gender differences in bicycling and motorcycling, Natural Language Processing (NLP) methods to understand the news, and research on semantic analysis. Specifically, we introduce the methods we used to develop accurate classifiers. Furthermore, we discuss the broader social aspects related to cycling.

\vspace{1mm} \noindent \textbf{Bicycling, Transportation, and Infrastructure Research.}
There has been substantial research into bicycling infrastructure and the public's perception of bicyclists using various methods (e.g., surveys, crowd-sourcing, and content analysis). \citet{macmillan2014societal} and \citet{macmillan2017understanding} used dynamics modeling to illustrate a feedback loop between political will, environmental views, and bicycling growth. ~\citet{aldred2019barriers} found that media and public opposition impact funding for new bicycling infrastructure in cities considering it. \citet{ferster2021advocacy} showed cyclical changes in attitudes toward bicycling infrastructure, while \citet{barajas2021biking} noted the lack of bike infrastructure in Black and Latino neighborhoods, suggesting that analyzing local news perceptions can help advocate for improved infrastructure in these communities. \citet{boufous2016reporting} highlighted that newspaper reports of cycling accidents often focus on dramatic, unusual incidents, contrasting with public health reporting and potentially misleading about cycling risks. \citet{macmillan2016trends} found a significant increase in media coverage of cyclist fatalities over two decades, indicating a specific media bias towards cyclists, unlike the steady coverage of motorbike accidents.

\vspace{1mm} \noindent \textbf{Gender and Bicycling.}
Previous research~\cite{emond2009explaining} shows that female and male cyclists perceive safety and needs differently. This means that the impact of social media on their attitudes and behavior may also differ. Female cyclists are often more risk-averse and influenced by content addressing their specific concerns, such as the preference for off-road paths away from traffic~\cite{garrard2008promoting}. In contrast, male cyclists may respond more to news that aligns with their attitudes and perceptions, including higher exposure to severe crashes. \citet{singleton2016cycling} found that gender-based differences in attitudes, preferences, and social norms, including financial or logistical barriers, affect women's likelihood to cycle, suggesting a need for policy reevaluation to support female cyclists. \citet{prati2019gender} and \citet{bouaoun2015road} further report that women experience higher discomfort in mixed traffic and have a lower risk of crashes, whereas men, though facing higher injury death rates, have a slightly lower risk per trip. \citet{aitbihiouali2022inclusive} showed that dedicated cycling infrastructure can increase female cycling participation by 4-6\%. These findings highlight the importance of understanding the demographic determinants of cyclists' safety perceptions and the need for tailored approaches to bridge gender gaps in cycling~\cite{emond2009explaining, garrard2008promoting}.

\vspace{1mm} \noindent \textbf{Linguistic Analysis.}
Drawing from established methods in prior studies~\cite{chen2014understanding, rho2018fostering}, we identified both content features like n-grams and stylistic features using the Linguistic Inquiry and Word Count (LIWC) dictionary~\cite{pennebaker2015development}. This LIWC analysis measured the frequency of tokens related to cognition, perception, and various linguistic styles, including emotional and temporal language. Our goal was to understand how these different features in headlines might affect reader perceptions, in line with research showing that the linguistic style and sentiment of writing significantly influence audience reactions on news platforms and social media~\cite{beasley2015emotional}. Additionally, Media agents often employ different headlines with partisan biases for the same event, even when the article content is almost identical~\cite{silverman2016hyperpartisan,rho2018fostering}. Studies have also shown that people often share articles based on the headline alone, without fully reading the story~\cite{silverman2016hyperpartisan}. This indicates a common behavior of circulating news based on headlines, which suggests that the way headlines are perceived can significantly influence individuals' decisions and opinions.

\vspace{1mm} \noindent \textbf{Semantic Analysis and Understanding the News.} 
Semantic analysis in NLP has led to extensive research on text understanding, such as Frame Semantics in FrameNet~\cite{baker1998berkeley,fillmore2003background}, which interprets different perspectives of the same event. While \citet{card2015media} developed a media frames corpus for news article annotation, it lacked insights into specific entity perceptions. Recent work by \citet{sap2020social} has delved into social bias using frame semantics, combining pragmatic inference with commonsense reasoning. Alongside this, new tools have emerged, such as \citet{liu2021visual} multi-modal feature fusion method for image captioning, \citet{spinde2021neural} dataset for detecting biased news, and \citet{ang-lim-2022-guided} model for analyzing inter-organization relationships from news content, providing enhanced support for news writers. Most similar to our work is the study by ~\citet{gabriel-etal-2022-misinfo} that creates ``perceived'' news frames from headlines regarding how people think about misinformation, focusing on how readers interpret the intent behind the news. This approach parallels \citet{tourni2021detecting} development of the Gun Violence Frame Corpus (GVFC), where different news articles are analyzed for varied perspectives on the same topic. However, our work differs in two direct ways: first, it focuses specifically on the portrayal of bicycling in news headlines, analyzing implicit sentiments toward cyclists; and second, unlike previous studies that mainly examined direct textual categories~\cite{tourni2021detecting}, our approach delves into implicit sentiment towards the cyclist mentioned in the headline from the writer's perspective, offering a deeper insight into media representation of cycling.

\section{DATA COLLECTION AND ANNOTATION}


\vspace{1mm} \noindent \textbf{Data Collection.}
We use Google News to collect cyclists' and accident-related news headlines from 2001 to 2021. Google News is a popular news aggregation service that has been available on Google for a long time. It aggregates the latest news across multiple sources and categorizes them based on trends, regions, and topics. We use Google News XML RSS Feed API~\footnote{\url{https://news.google.com/news/rss.}}. The API limits the number of articles we can receive each month. We use the keywords ``cycling'', ``cyclist'', and ``bike'' to pull headlines. We scraped 32,980 for cyclist-related headlines. For annotation, we randomly sample 50\% of the headlines with the keywords ``crash'', ``death'', ``killed'', and ``dead'' and 50\% without these keywords. Annotators are uniformly provided a random sample generated from the entire dataset and the sub-sampled subset. We use the second set of keywords to guarantee that accident-related posts are annotated. However, it is important to note that accident-related posts can also appear without the sub-sampled keywords. To identify US-based headlines, we manually reviewed the URLs of news agencies from the data. This ensured that our selection was based on the news agency's location rather than the headline content.

\vspace{1mm} \noindent \textbf{Annotation Guidelines.}
We annotate two major categories for each news headline: Perception and Accident categories. Each category helps understand how people interpret the writer's perception of the cyclists in the headline, whether the article is about an accident, and the readers perceived opinion about who caused the accident.

\vspace{1mm} \noindent \textbf{Perception Categories.}
Perception is used to analyze and measure how readers perceive how readers interpret the writer's perception towards the cyclists in the headline. Specifically, we ask annotators to label perception into Negative, Positive, and Neutral categories.

\vspace{1mm} \noindent \textbf{\textit{Negative}.} The negative class is used if the reader perceives the writer's attitude towards the cyclist as negative. Negative perceptions can manifest in various ways. For example, the headline

\begin{quote} 
        {\textbf{Example}: Cyclist harasses motorists at Serangoon roundabout, smacks vehicles while hurling vulgarities---The~Independent}
\end{quote}
provides an example where a cyclist is claimed to have caused damage to vehicles. Likewise, negative perception can be caused because of explicit mentions of breaking traffic laws, for example, the headline
\begin{quote}  
        \textbf{Example:} Dashcam sees bus \& cyclist ignore red light in Norwich - Norwich Evening News
\end{quote}
discusses a cyclist running a red light. Generally, annotators agreed that it is negative when a cyclist is portrayed as committing a traffic crime. Upon inspection of the article, someone submitted a video because of the danger that can be caused by a bus running a red light, which can be perceived as a much greater danger than a cyclist running a red light. However, the writer found it equally important to mention the cyclists because their appearance in the video was interesting and relevant to the writer's implicit viewpoint. Thus, we argue that the writer is implicitly reporting the cyclist negatively by equating their actions with the more severe actions of the bus driver.

\vspace{1mm} \noindent \textbf{\textit{Positive}.} A positive perception reflects the reader's interpretation of the writer's headline as being positive about the cyclist. This can manifest in multiple ways. For example, the headline
\begin{quote} 
        \textbf{Example:} A Plover man donated a kidney to save a stranger. Now he's cycling cross-country to raise awareness. - Stevens Point Journal
\end{quote}
discusses how a cyclist is raising awareness for organ donation. On the other hand, there are many examples where an accident is discussed where the cyclist describes direct mention of severe injuries. For instance, the headline
\begin{quote} 
        \textbf{Example:} Speeding taxi drags cyclist to his death, leaves another injured - DispatchLIVE
\end{quote}
describes a cyclist being killed in an accident. In such gruesome (i.e., descriptive) descriptions, annotators assumed that the writer expressed a negative sentiment towards the driver. Yet, the writer's perception of the cyclist is positive, i.e., the writer would not write such a descriptive narrative of the accident if they thought negatively about the cyclist. Hence, even though the writer reports a gruesome event, it is perceived to raise awareness about the driver's actions. Less descriptive headlines were labeled as neutral. This is the case in prior work by \citet{joye2015domesticating} where establishing an emotional bond (e.g., ``panic and chaos everywhere'') can be a tactic to get the audience to care. Finally, in 
\begin{quote} 
        \textbf{Example:} Remorseful taxi driver keeps license after cyclist crash - alloaadvertiser.com
\end{quote}

\vspace{1mm} \noindent \textbf{\textit{Neutral}.} There are many headlines where it is unclear whether the writer views the cyclists with positive or negative perceptions. We label such headlines as Neutral. For example, the headline
\begin{quote} 
        \textbf{Example:} French cyclist Brunel signs two-year deal with UAE Team Emirates - Gulf Today
\end{quote}
provides a simple fact about an athlete. The text has little emotional expression, and it is impossible to determine whether the writer is a fan of the athlete. On the other hand, the headline
\begin{quote} 
        \textbf{Example:} SH 183 reopened in Cedar Park after cyclist-vehicle crash - KXAN.com
\end{quote}
reports a crash similar to the positive headline discussed above. However, the description of this headline is not directly about the accident---it is focused on the reopening of a highway. Hence, it is difficult to determine how the writer feels about the cyclist.

\begin{table}[t]
\centering
\resizebox{.65\linewidth}{!}{
\begin{tabular}{lll}
\toprule
& \textbf{Category} & \textbf{Cyclists} \\
\midrule
\multirow{2}{*}{\textbf{Related}}  & Yes  & 623 \\
                                   & No   & 877 \\
\midrule
\multirow{3}{*}{\textbf{Fault}}    & Cyclist  & 36 \\
                                   & Unknown  & 1163 \\
                                   & Other    & 301 \\
\midrule
\multirow{3}{*}{\textbf{Perception}} & Negative & 94 \\
                                     & Neutral  & 236 \\
                                     & Positive & 1170 \\
\bottomrule
\end{tabular}
}
\caption{Dataset Statistics for the labeled data.\vspace{-1em}}
\label{tab:stats}
\end{table}


\vspace{1mm} \noindent \textbf{Accident-Related Categories.} We annotate headlines into two major classes: Related to an Accident (Yes/No) and Accident Fault (Cyclists, Unknown, Driver). 

\vspace{1mm} \noindent \textbf{\textit{Related to Accident (Yes/No)}.} A headline is related to an accident if it is directly discussed in the headline. For instance, the headline
\begin{quote}
        \textbf{Example:} Caught On Camera: Passenger Opens SUV Door, Seriously Injuring Cyclist in Lincoln Park
\end{quote}
directly mentions a crash involving cyclists (we use the same criteria for motorcyclists). Likewise, we annotate headlines that involve an ``attack'' of some kind (e.g., a fight, assault, etc.) as being related to an accident. For example, the headline
\begin{quote}
        \textbf{Example:} Cyclist attacked by two unknown tracksuited men on Boothferry Road --- Hull Live
\end{quote}
mentions cyclists actually getting attacked by two track-suited men.

The ``Not Related'' category is used for headlines that do not mention anything related to an accident. For instance, the headline
\begin{quote}
        \textbf{Example:} Gallery: A Paris-Roubaix for the ages --- Cyclist
\end{quote}
discusses a race, not an accident or attack.

\vspace{1mm} \noindent \textbf{\textit{Fault}.}
If a headline is related to an accident, we annotate it with who is perceived to be at fault by readers, i.e, who caused the accident. Specifically, annotators will label each headline as ``Cyclists'', ``Unknown'' or ``Other''. For instance, the ``\textbf{Cyclists}'' fault headline
\begin{quote} 
        \textbf{Example:} Blame entirely on the cyclist. Cyclist collided with truck
\end{quote}
directly puts the blame \textit{entirely} on the cyclist; hence it is labeled as being the cyclist's fault.
As an example of the ``\textbf{Unknown}'' class, the headline
\begin{quote}
        \textbf{Example:} SH 183 reopened in Cedar Park after cyclist-vehicle crash  - KXAN.com
\end{quote}
shows a fact and does not provide details about who caused the accident. Finally, for the ``\textbf{Other}'' class, in the headline
\begin{quote}
        \textbf{Example:} Alludes to hit and run on the vehicle owner’s part.
\end{quote}
we see an example where a driver was involved in a hit-and-run. In this case, someone ``Other'' (i.e., the driver) is at fault.

\vspace{1mm} \noindent \textbf{Agreement and Dataset Statistics.}

\begin{table}[t]
\centering
\resizebox{0.8\linewidth}{!}{%
\begin{tabular}{llll}
\toprule
          & \textbf{Related} & \textbf{Fault} & \textbf{Perception} \\ \midrule
Cohen's Kappa & .90 & .71 & .63 \\ \bottomrule
\end{tabular}%
}
\caption{Cohen's Kappa between the two annotators for each category for all of our labeled data.\vspace{-1em}} 
\label{tab:agreement}
\end{table}

We followed a two-stage annotation procedure. First, two annotators completed the process independently of one another. One-on-one interviews were conducted with annotators to identify dataset issues versus annotator errors. The information gained from the discussions was then used to improve the guidelines. Next, the two annotators completed 1,500 headlines related to cyclists. The agreement results are shown in Table~\ref{tab:agreement}. The Cohen's Kappa agreement scores ranged from .62 to .90, representing ``substantial agreement'' and ``almost perfect agreement,'' respectively~\cite{landis1977measurement}. 
To improve the accuracy and consistency of perception annotations, we introduced a third annotator to help adjudicate areas with lower agreement scores. This step significantly improved the final dataset, as evidenced by a Cohen's Kappa of 0.63 for Perception, indicating a substantial agreement (where 1 indicates perfect agreement). Second, to improve data annotation further, we again met with the annotators and reviewed and discussed disagreements among the annotated headlines to form the final gold standard dataset. The final dataset statistics can be found in Table~\ref{tab:stats}.


\vspace{1mm} \noindent \textbf{US-Based Website Dataset Statistics.} \textcolor{black}{In our dataset, we have a total of 31,480 non-labeled news headlines and 1,500 labeled ones. These headlines are collected from a total of 5,827 unique website links worldwide. Among these, 2,168 are based in the US, and within the US-based websites, there are 125 domain-specific links and 2,043 general news website links. In our work, we manually identified all of these website links by reviewing each link one-by-one. Geographical factors play a significant role in cycling activities. Literature suggests that geographical contexts significantly influence cycling activities~\cite{relia2018socio}. For instance, factors such as infrastructure, climate, and traffic regulations in the US might differ from other regions~\cite{chan2020climate}, thereby affecting cycling prevalence and the nature of cycling-related incidents reported in the media. Hence, our analysis is narrowed down to US-based websites. Based on these websites, we refine our data by filtering news headlines based on pronouns such as 'he' or 'she' to categorize them into male and female-related headlines. The statistics derived from this categorization are illustrated in Table~\ref{tab:stats2}.}


\begin{table}[t]
\centering
\resizebox{.55\linewidth}{!}{
\begin{tabular}{lr}
\toprule
\textbf{Geographic Region} & \textbf{Cyclists} \\ \midrule
Non US-Based & 20,095 \\
US-Based & 11,385 \\ 
\bottomrule
\end{tabular}}
\caption{Dataset Statistics for the Non-labeled Data.\vspace{-1em}}
\label{tab:stats2}
\end{table}


\vspace{1mm} \noindent \textbf{FAIR Requirements.}
Our dataset adheres to FAIR principles (Findable, Accessible, Interoperable, and Re-usable). The dataset \textit{will be} Findable and Accessible through Zenodo~\footnote{The dataset is in the supplementary material for now.}. Moreover, the data will be licensed under the Creative Commons Attribution License (CC BY 4.0). Finally, the data is shared as a text file format along with the complete annotation guidelines, which are shared as Word documents. Thus, the data is reusable and interoperable.

\section{METHODOLOGY}



We use large language models and prompting to classify cyclist perception Previous studies on enhancing large language models' reasoning have mainly focused on two methods. One method involves generating rationales in natural language, such as Chain-of-Thought~\cite{wei2022chain}, or code, like Program-of-Thought~\cite{chen2023program}. These approaches execute reasoning step-by-step in real time without a separate planning phase. As a result, models must process and apply logic in a single operation, limiting their expressiveness. Moreover, these models cannot reuse previously understood logic when encountering similar problems, impacting their efficiency~\cite{chae2024language}.

The second strategy to improve LLMs is by generating detailed plans in natural language, which are then broken down into specific reasoning steps, seen in methods like Least-to-Most~\cite{zhou2022teaching} and Plan-and-Solve~\cite{wang2023plan}. However, prior researches suggest that natural language may not be the most effective way to convey task logic. As a solution, ~\citet{li2023chain} introduced Chain-of-Code (CoC), which uses pseudocode to manage undefined behaviors more efficiently.  This allows models to simulate expected outputs and manage unexecutable code lines, improving their reasoning abilities. Additionally, traditional methods focus on single instances, which overlooks the potential to identify and use common reasoning patterns across similar tasks~\cite{zhou2024self}. \citet{chae2024language} addressed this with the THINK-and-EXECUTE method, which identifies and applies these common patterns to improve reasoning in LLMs.

Inspired by \citet{li2023chain} and \citet{chae2024language}, we use programming languages like Python to formulate the logic needed to address specific tasks, such as those in the PoT or CoC, taking advantage of their strict and precise syntax. We opt for pseudocode over natural language to more effectively guide the reasoning processes of language models, which are typically trained on natural language instructions. Our approach breaks down complex problems into a series of intermediate reasoning steps. This method streamlines the development process by prompting for each step in the sequence. Our logic's execution is simulated by an LLM. Our approach is structured into two primary steps (as shown in Figure~\ref{fig:bikeframe}): (1) \textbf{DESIGN}, where we develop code to address each sub-problem associated with the Bike Frame tasks methodically, and (2) \textbf{EXECUTE}, where this code is implemented by invoking the ``BikeFrame'' class. Here, we input test headline titles and news sources as shown in Figure~\ref{fig:bikeframe} within the LLMs block.  We describe each subsection next. 

\subsection{DESIGN: Pseudocode}

\begin{figure}[t]
    \centering
    \includegraphics[width=1\linewidth]{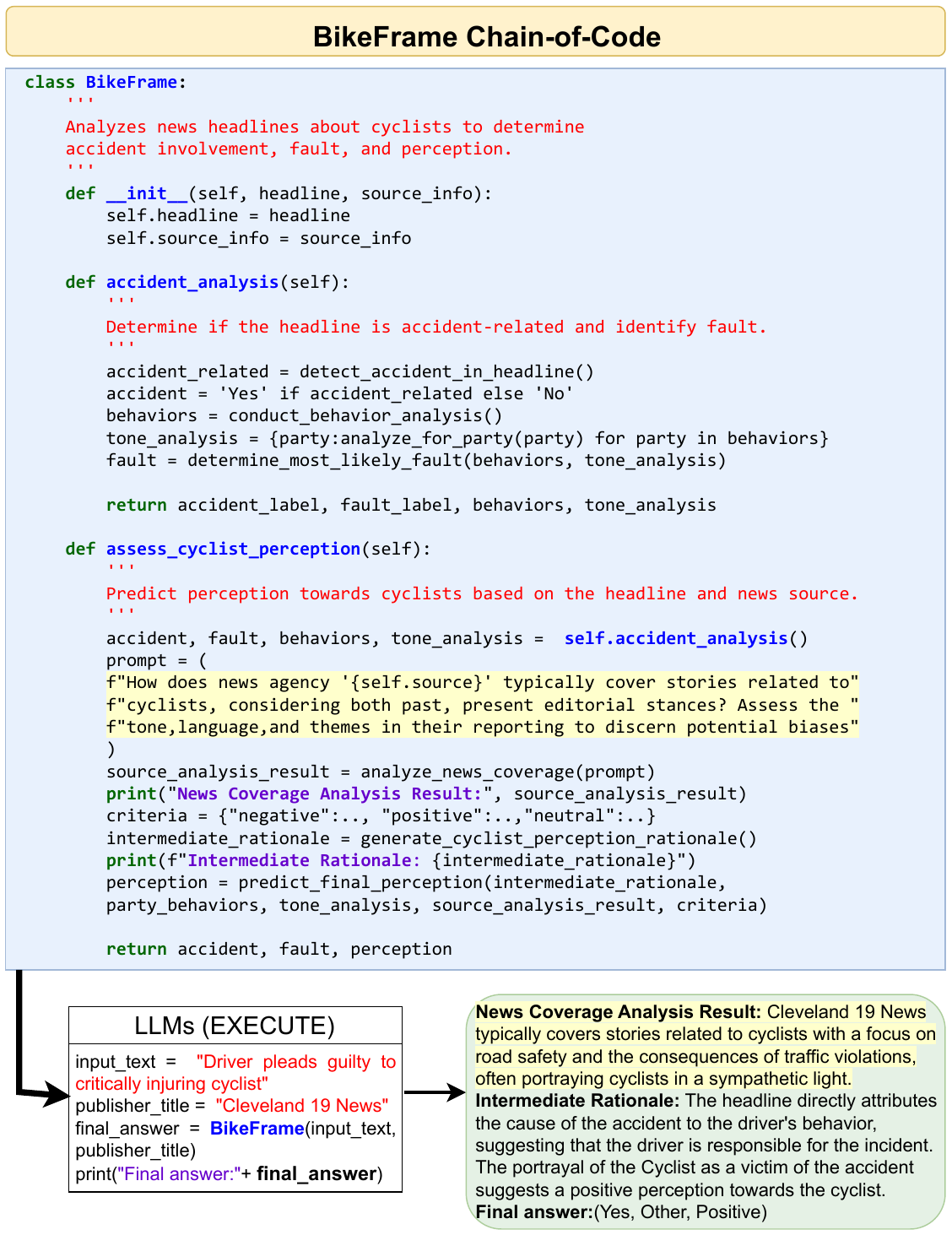}
    \caption{BikeFrame Chain-of-Code prompt. The news information component is highlighted in \hl{yellow}.} \vspace{-1em} 
    \label{fig:bikeframe}
\end{figure}

We developed a Chain-of-Code prompt named the ``BikeFrame,'' which jointly predicts three tasks: determining if an accident happened, identifying who is at fault, and assessing the perception toward cyclists. The ``BikeFrame'' class in the prompt includes two main functions for accident detection and perception analysis, shown in Figure~\ref{fig:bikeframe}. First, the accident analysis function checks the headline for mentioning accidents (e.g., collisions or injuries). The class description, highlighted in red in Figure~\ref{fig:bikeframe}, guides the LLMs in performing this analysis. If an accident is detected, the function conducts a behavior analysis to identify actions or failures contributing to the incident, extracting behaviors of all parties involved and assessing if traffic laws were violated. The class then evaluates the tone of the headline to determine the perception towards each party. The results from the accident analysis are passed to the cyclist perception function via a return statement. In the perception function, these analyses are combined to identify the most likely party at fault and predict the overall perception towards cyclists. This pseudocode structure makes our prompt flexible, allowing it to incorporate different reasoning components.

\begin{table*}[t]
\centering
\resizebox{.95\linewidth}{!}{%
\begin{tabular}{lrrrrrrrr}
\toprule
                                                           & \multicolumn{3}{c}{\textbf{Perception Categories}} & \multicolumn{4}{c}{\textbf{Accident Categories}}                & \multicolumn{1}{l}{} \\ \cmidrule(lr){2-4} \cmidrule(lr){5-8}
                                                           & \textbf{Negative}    & \textbf{Neutral}   & \textbf{Positive}   & \textbf{Not Accident} & \textbf{Cyclist} & \textbf{Unknown} & \textbf{Other} & \textbf{Avg}                             \\ \midrule
\textbf{Uniform}                                           & .094                 & .176               & .470                & .513                  & .018             & .415             & .241           & .275                                         \\
\textbf{Stratified}                                        & .051                 & .093               & .793                & .561                  & .000             & .761             & .286           & .364                                         \\ \midrule \midrule
\textbf{Logistic Regression}                               & .338                 & .488               & .829                & .911                  & .154             & .930             & .811           & .637                                         \\
\textbf{Logistic Regression + LIWC}                        & .174                 & .514               & .899                & .908                  & .182             & .949             & .847           & .639                                         \\ \midrule
\textbf{RoBERTa}                                           & .444                 & \textbf{.714}      & .935                & .962                  & .105             & .942             & .853           & .708                                         \\
\textbf{MT RoBERTa}                                        & .710                 & .696               & .940                & .960                  & .154             & .957             & .880           & .757                                         \\
\textbf{MTLPT RoBERTa}                                     & .519                 & .677               & .933                & \textbf{.970}         & .133             & .959             & .898           & .727                                         \\
\textbf{MT + MTLPT RoBERTa}                                & .583                 & .646               & .936                & \textbf{.970}         & .200             & .959             & .893           & .741                                         \\ \midrule
\textbf{Zero-Shot GPT3.5}                                  & .238                 & .372               & .510                & .868                  & .364             & .853             & .671           & .554                                         \\
\textbf{Zero-Shot GPT3.5 + CoT}                            & .320                 & .309               & .534                & .841                  & .500             & .819             & .635           & .565                                         \\ 
\textbf{Few-Shot GPT3.5}                                   & .253                 & .344               & .433                & .918                  & .421             & .898             & .809           & .582                                         \\
\textbf{Few-Shot GPT3.5 + CoT}                             & .282                 & .438               & .723                & .898                  & .471             & .897             & .655           & .624                                         \\ \midrule 
& \multicolumn{8}{c}{\textbf{Our Methods}}   \\ \midrule
\textbf{Zero-Shot GPT3.5 + CoC  + News}                    & .119                 & .395               & .406                & .787                  & .094             & .669             & .479           & .421                                         \\
\textbf{Zero-Shot GPT3.5 + CoC  + News + Self-Consistency} & .151                 & .516               & .394                & .778                  & .094             & .652             & .480           & .438                                         \\ 
\textbf{Few-Shot GPT3.5  + CoC  + News + Self-Consistency} & .759                 & .684               & \textbf{.947}       & .941                  & \textbf{.667}    & \textbf{.956}    & \textbf{.903}  & \textbf{.837}                                \\ \midrule
& \multicolumn{8}{c}{\textbf{Ablation}}   \\ \midrule
\textbf{Few-Shot GPT3.5  + CoC  + News}                     & \textbf{.813}        & .667               & .938                & .931                  & .545             & .943             & .872           & .815                                         \\
\textbf{Few-Shot GPT3.5  + CoC  + Self-Consistency}      & .583                 & .417               & .776                & .934                  & .600             & \textbf{.956}    & \textbf{.904}  & .739                                         \\ 
\textbf{Few-Shot GPT3.5  + CoC }                           & .296                 & .402               & .759                & .927                  & .545             & .947             & .886           & .680                                         \\
 \bottomrule
\end{tabular}%
}
\caption{F1 scores for all classes. The best scores are \textbf{bolded} for each column. \vspace{-1em}}\label{tab:joint-preds}
\end{table*}

Additionally, A novel feature of our BikeFrame class is the \textbf{\textit{integration of an analysis of the news agency's reporting style}} by evaluating the tone, language, and themes in their coverage of cyclist-related stories. This analysis helps discern any biases or recurring patterns in the agency's reporting. By combining these insights, the LLMs identify the most likely party at fault and predict the overall perception towards cyclists. This method ensures a comprehensive understanding of how the news agency's style influences public perception, supported by studies that emphasize the importance of media framing in shaping public opinion~\cite{sharof2024exploring}. We simply add a prompt in the pseudocode to extract the opinion of the news source from LLMs based on their pretrained knowledge. To ensure the LLMs use this news source analysis in subsequent decisions, we use intermediate \texttt{print()} statements that explicitly reference the analysis. These statements clarify the rationale behind each decision, illustrating how the model arrives at its conclusions. The sections where we integrate the news source analysis are in yellow in Figure~\ref{fig:bikeframe}. 



\subsection{EXECUTE: Simulate Pseudocode Execution}

The execution process is visually demonstrated in a green box in Figure~\ref{fig:bikeframe}, representing the print outputs for news coverage analysis, intermediate rationales, and the final answer for joint classification tasks. This piece of the prompt is appended to the design prompt, representing an instantiation of the class before passing to the LLM. By employing the \texttt{print()} statement, we ensure that the LLMs strictly adhere to the pre-established rules of our Bike Frame. This structured approach enhances the model's adherence to logical constraints and substantially reduces errors and hallucinations, thus improving the reliability and accuracy of the model's output. Prior studies have shown that displaying intermediate steps is vital for logically reaching the final answer through CoT rationales~\cite{wei2022chain, chae2024language}. Monitoring these intermediate steps helps the models maintain an accurate state of variable changes throughout the code execution. Following the execution prompt design of ~\citet{chae2024language}, the final output for joint tasks is displayed using \texttt{print(Final answer: \{final\_answer\})} as the last output of the system.

\section{RESULTS AND DISCUSSION}
This section describes our experimental setup, performance results, and a use study using our models on all US-based news headlines, both labeled and unlabeled. Specifically, we analyze reporting differences across cyclist articles, general and domain-specific websites (like bicycling.com), and when gender-related names are mentioned. Additionally, we study linguistic factors in headlines, including style and content, to understand their impact on perceptions of cyclists, both positive/negative and being at-fault in accidents.

\vspace{1mm} \noindent \textbf{Experimental Setup.}
We employ GPT-3.5-Turbo~\cite{OpenAI_chatgpt} for its strong reasoning and code generation capabilities due to its balance of performance and cost-effectiveness. We set the temperature parameter T=0.0 to enable greedy decoding, ensuring outputs are precise and deterministic. To further enhance the model's performance and robustness, we adopt the Self-Consistency approach~\cite{wang2022self}, generating multiple outputs using the chain-of-code framework and aggregating them to derive the final answer. Following the setting in \citet{wang2022self}, we use a top-k sampling set at 0.9 with 20 sampled reasoning paths to refine our decision-making process, maintaining a reasonable cost-to-performance ratio. We trained multiple baseline models on our dataset, splitting it into train, dev, and test sets with a ratio of 7:1:2, using four NVidia GeForce GTX 1080 Ti GPUs. Each model was trained on the training set and tested to calculate the F1 score for each class. 

\vspace{1mm} \noindent \textbf{Baselines}
We evaluate eight baseline models: Logistic Regression with Tfidf ngram features (LR), Logistic Regression with Tfidf ngram and LIWC features, three RoBERTa variant models with multiclass (multiclass, multi-task, multi-task MTLPT (see appendix for details), and two random baselines. We also evaluate additional prompt-based baselines, including (1) Direct prompting, where the model predicts answers without generating rationales (2) Zero-shot CoT~\cite{kojima2022large}, where add ``Let think step-by-step'' to prompt LLM solving problems through sequential reasoning. (3) Few-Shot CoT~\cite{brown2020language}, which uses a few examples to enhance its reasoning.

\vspace{1mm} \noindent \textbf{Model Performance.} In Table~\ref{tab:joint-preds} we report the F1 score for each of the Accident-related classes. Our results show that the RoBERTa model significantly outperforms all Linear and Random baselines, as well as our BikeFrame Chain-of-Code model (.970 vs .941) in detecting accident-related content. This is likely because detecting accident patterns is a simple task, making it easier for RoBERTa to identify the relevant features. However, for the ``who is at fault'' and perception tasks, our BikeFrame method with news source analysis and self-consistency outperforms all RoBERTa variant models in every class except for the ``Neutral'' classes. When comparing Few-Shot GPT-3.5 + CoC + News + Self-Consistency to the Zero-Shot method, we find that it significantly improves performance across all classes. The largest improvements are seen in the ``Negative'' class, which involves complex reasoning chains. To accurately decide on perception, it is often necessary to consider comprehensive situations such as whether the cyclist violated traffic laws and how the cyclist is described using sarcasm or passive voice. For the "Cyclist at fault" class prediction, our Few-Shot GPT-3.5 + CoC + News + Self-Consistency model outperforms the best RoBERTa variant, "MT + MTLPT RoBERTa." This class is a low-resource category with only 94 cases, much fewer than other categories. Our model effectively learns patterns from few examples, significantly enhancing performance, which allows the model to better generalize from the small number of available instances. 

When comparing our CoC method with regular CoT and direct prompting, we find that for zero-shot learning, both CoT and direct prompting outperform our method. This may be because our prompts are more complex and, without examples, can make it difficult for language models to understand the underlying reasoning chain. However, we also observed that the output of our zero-shot learning approach adheres more closely to the desired output format. Our method results in significantly fewer parsing errors, whereas general CoT and direct prompting often produce outputs that do not fit our desired format or miss some predictions. Despite being zero-shot, our method successfully addresses these issues. Therefore, adding a few examples significantly improves the model's reasoning ability. Overall, our Few-Shot GPT-3.5 + CoC + News + Self-Consistency model outperforms all other models based on average F1 scores.


\begin{figure}[t]
    \centering
    \includegraphics[width=.65\linewidth]{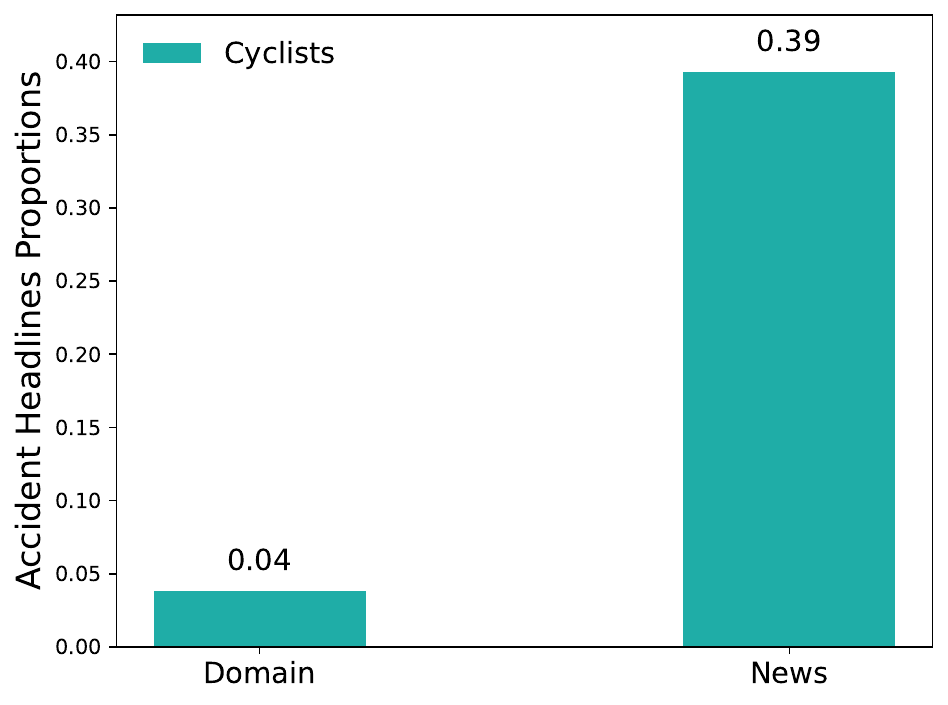}
    \caption{The proportion of accident-related headlines in the domain-specific (e.g., bicycling.com) and general news websites (e.g., nypost.com and chicagotribune.com).\vspace{-1em} } 
    \label{fig:groups}
\end{figure}

\vspace{1mm} \noindent \textbf{Ablation Studies}
We conduct an ablation study on each component of our CoC BikeFrame prompt. To do this, we prepare three types of pseudocode prompts: one without the news source analysis component, one without the self-consistency component, and one without both components. Table~\ref{tab:joint-preds} shows that the model without the self-consistency component slightly underperforms compared to our best model based on the average F1 score. The most significant performance drop occurs when the news source analysis component is removed, resulting in almost a 10\% reduction in the average F1 score. This drop is particularly noticeable in perception prediction, where the model without the news component performs significantly worse across all three classes (for negative .759 vs .583; for neutral .684 vs .417; and for positive .947 vs .776). These results indicate that incorporating comprehensive news source analysis significantly enhances our model's performance, especially in accurately predicting perceptions. This improvement is likely because the news source analysis helps the model understand how different news agents frame cyclists. By examining the language and context used in news articles, the model can better identify biases, sentiments, and the overall narrative presented about cyclists. This deeper understanding enables the model to make more accurate predictions regarding perceptions, as it can account for the subtle nuances and framing techniques used by various news sources.

\vspace{1mm} \noindent \textbf{RQ1: Are there differences in how domain-specific websites and general news websites attribute fault for accidents, specifically to cyclists?}
We applied our best model to analyze 31,480 US-based news headlines, selected from a larger dataset of 81,746 global headlines. Our study counted accident-related headlines in  cycling categories and classified 2,168 US websites as either domain-specific (like bicycling.com) or general news websites (such as nytimes.com). Out of these, 125 were domain-specific and 2,043 were general news sites. We then calculated the proportion of accident-related headlines for both types of sites, as shown in Figure~\ref{fig:groups}. Results showed that general news sites are more likely to report accidents involving cyclists than domain-specific ones, with significant differences (p-value $<$ .00001 using Z-test). This pattern suggests that media consumption could influence perceptions of cyclist safety, aligning with studies on how safety perception affects driving behavior~\cite{intravia2017investigating}. These findings raise questions about the impact of media reporting on cycling behavior and suggest that changes in reporting could influence future cycling growth~\cite{alvisyahri2020motorcyclist}.

\begin{figure}[t]
    \centering
    \includegraphics[width=.7\linewidth]{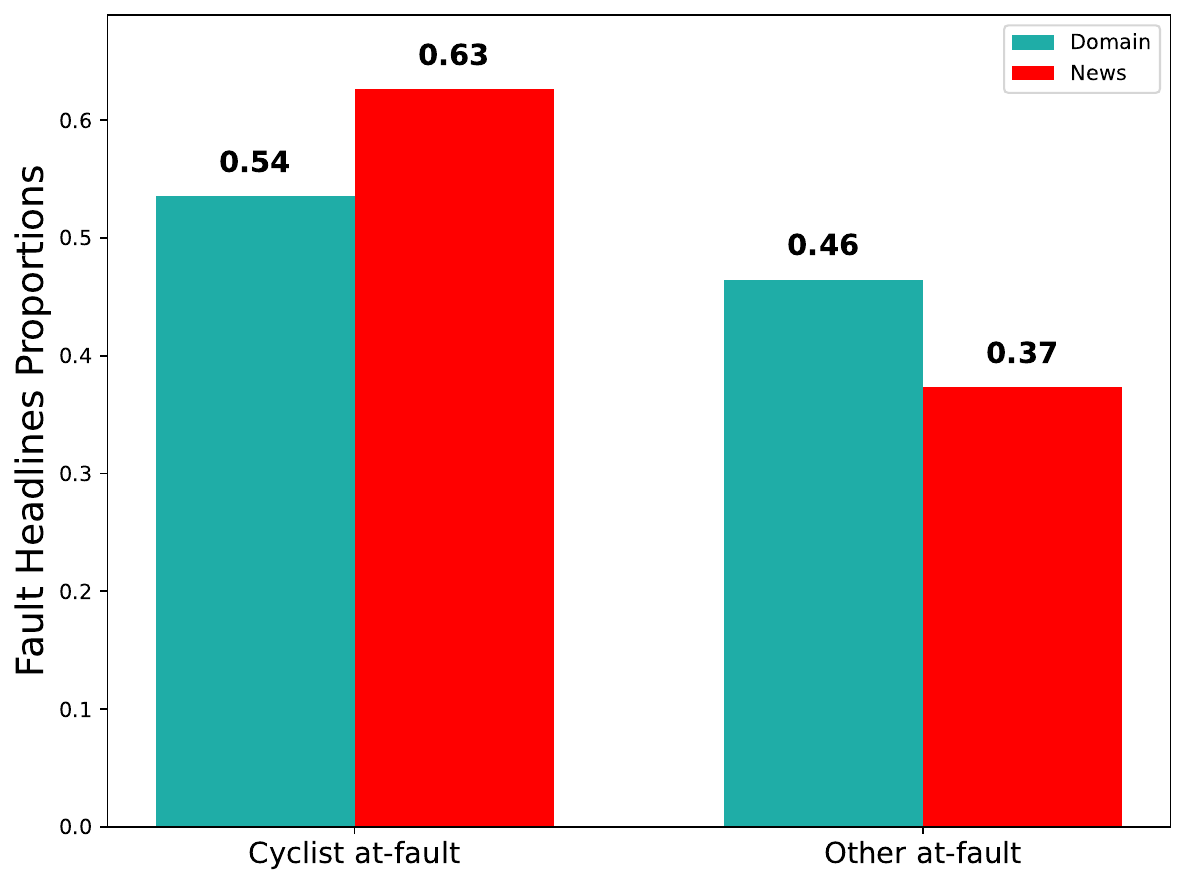} 
    \caption{The proportion of headlines in domain-specific websites vs. general news-based websites on who is perceived to be at-fault for the accident, the cyclists, or other entities in the headline (Other). \vspace{-1em}}
    \label{fig:groups2}
\end{figure}

In Figure~\ref{fig:groups2}, we report the proportion of headlines in the domain-specific and general news categories that state that whether cyclists or other entities are at fault for an accident. This figure answers the question, which websites are more likely to report negative stories about cyclists? General news websites make the highest proportion of at-fault stories. Specifically, 63\% of the headlines in general news websites attribute fault to cyclists, compared to 54\% in domain-specific websites. Conversely, 46\% of headlines in domain-specific websites attribute fault to other entities, compared to 37\% in general news websites. These findings suggest that general news websites are more likely to frame cyclists negatively in accident-related stories.

\begin{figure}[t]
    \centering
    \includegraphics[width=.7\linewidth]{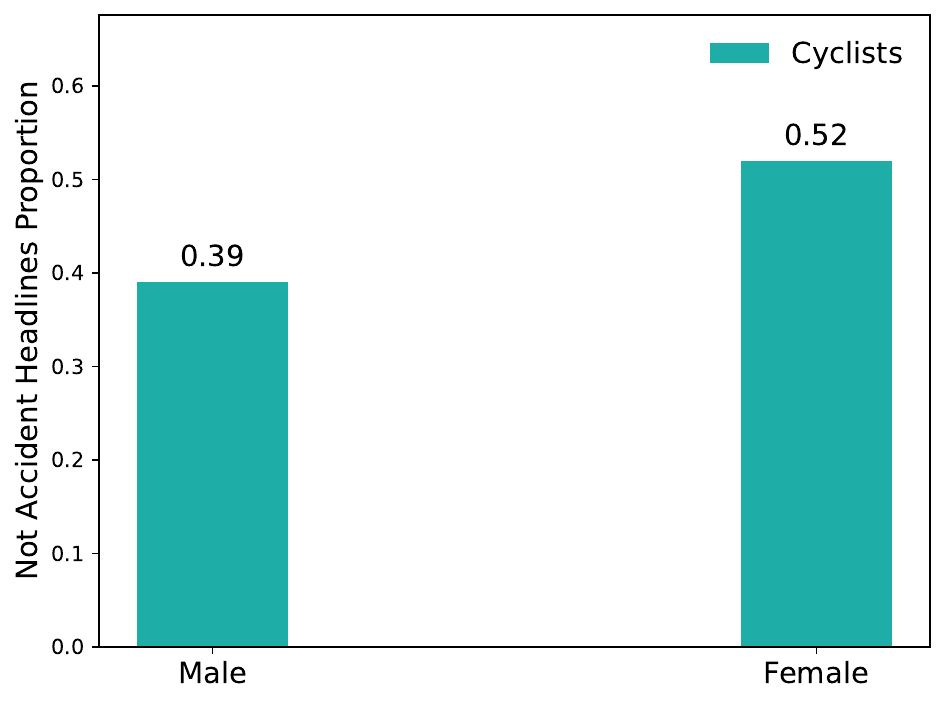}
    \caption{The proportion of male/female headlines unrelated to an accident for cyclists.\vspace{-1em}}
    \label{fig:not-accident}
\end{figure}

\vspace{1mm} \noindent \textbf{RQ2: Do news agencies report cycling accidents for males and females in proportion to their respective accident rates and presence in the cycling population?}
Our study investigates gender disparities in cycling-related news coverage. As mentioned, female cyclists are often influenced by news addressing specific needs, while male cyclists are influenced by news that aligns with their attitudes and perceptions~\cite{emond2009explaining}. Additionally, fatalities among women cyclists are often more prominently covered in the news headlines, possibly due to the perception of women as more vulnerable road users, making female-related accidents more newsworthy~\cite{mindell2012exposure}. 

Despite male cyclists have a significantly higher risk of accidents, with death rates six times and injury rates five times higher than females~\cite{centers2003web}, our analysis using gendered pronouns in news headlines reveals uneven media coverage. If accidents were reported equally in the media, we would expect significantly more headlines about male-related accidents. Yet, Figure~\ref{fig:not-accident} shows the proportion of non-accident-related headlines is lower for males than females in both cycling (.39 vs. .52), which is statistically significant (p-value $<$ .0001). Interestingly, the actual number of male-related accident reports are only 2.04 times higher than those for females, not aligning with the actual higher risk. If media coverage reflected actual accident rates, we would expect significantly more headlines about male-related accidents. However, our findings suggest that female cyclists receive disproportionately more coverage, indicating potential media bias. This bias highlights the need for policies and interventions to support and encourage women in cycling, considering their different safety perceptions, accident rates, and infrastructure preferences~\cite{prati2019gender,aitbihiouali2022inclusive,singleton2016cycling,bouaoun2015road}. This study contributes to a broader understanding of how gender stereotypes in media may not align with actual demographic patterns, emphasizing the importance of equitable representation in urban transportation.


\begin{figure}[t]
    \centering
    \includegraphics[width=.7\linewidth]{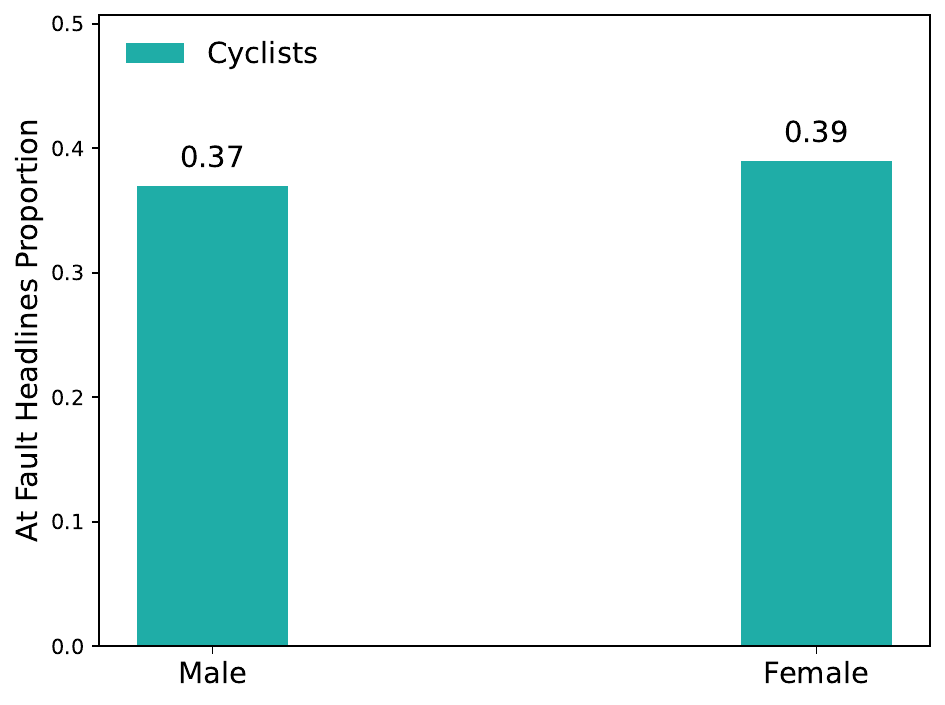} 
    \caption{The proportion of at-fault headlines for cyclists in the Male and Female groups.\vspace{-1em}}
    \label{fig:at-fault}
\end{figure}


Figure~\ref{fig:at-fault} shows that male cyclists are perceived as less at-fault in accidents than female cyclists (.37 vs .39), potentially due to stereotypes about female cyclists being novel riders. This is consistent with research showing that female drivers tend to be more cautious~\citet{al2003role}, while male riders are often seen as more skilled but prone to risky behaviors, such as driving under the influence of alcohol.  These findings emphasize how gender perceptions affect fault attribution in cycling accidents, underlining the importance of further research into these dynamics and their impact on road behaviors and safety.

\vspace{1mm} \noindent \textbf{RQ3: What linguistic factors in news headlines predict the perception towards cyclists?}
Psycholinguistic research highlights that language use in social media significantly affects emotions, behaviors, and perceptions~\cite{chen2014understanding,ernala2017linguistic,park2011politics}. The complexity of readers' reactions to news headlines involves cognitive aspects like inferring intent, emotional responses such as feelings of distrust, and behavioral actions like sharing news. These reactions are influenced by both the content and linguistic style of the headlines~\cite{gabriel-etal-2022-misinfo}. We used a machine learning classifier trained on 3500 manually annotated news headlines to gauge public perceptions of cyclists, incorporating the top 500 n-grams and 93 LIWC features. Our analysis identified the 20 most influential n-grams and LIWC features that shape perceptions of fault and general sentiment towards cyclists and motorcyclists. Specifically, our study delves into those that portray cyclists as at-fault with negative perceptions, as shown in Figure~\ref {fig:lex-temp}.



\begin{figure}[t]
    \centering
    \includegraphics[width=\linewidth]{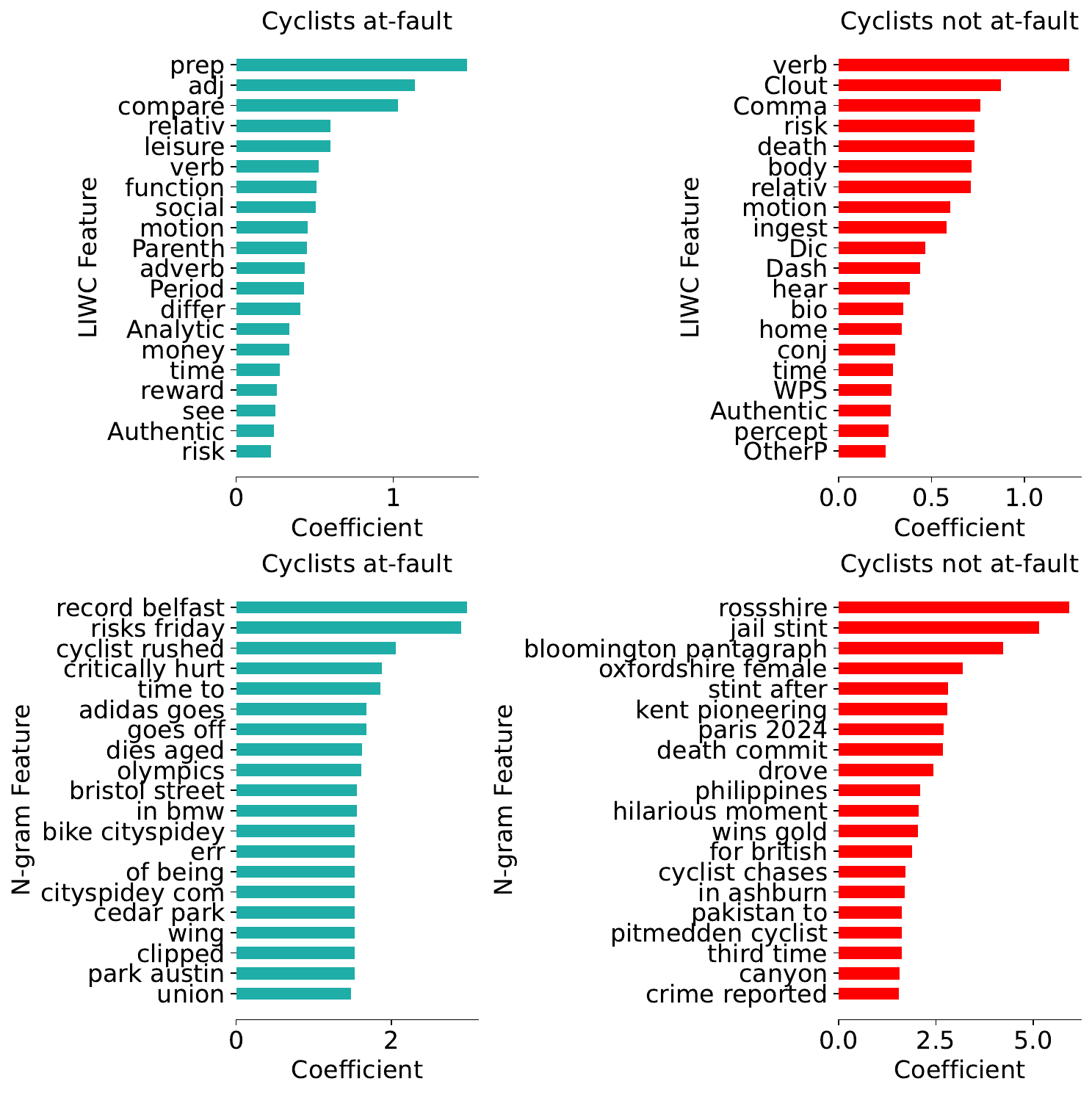}
    \caption{Top 20 Linguistic Factors (LIWC and n-grams (n=1,2)) in news headlines for cyclists at-fault or not.\vspace{-1em}}
    \label{fig:lex-n-gram-at_fault}
\end{figure}

\vspace{1mm} \noindent \textbf{Linguistic Style for Cyclists At-Fault.} In Figure~\ref{fig:lex-n-gram-at_fault}, we can see that headlines shape public perceptions of at-fault cyclists using language that elicits strong emotions and cognitive evaluations, with features like  like ``prep,'' ``compare,'' ``leisure,'' ``relate,'' and ``affective.'' These elements suggest both an emotional response and a critical assessment of cyclists' actions, which can bias reader perceptions and contribute to negative sentiments on social media~\cite{liao2014can}. For instance, the use of words like "prep" and "compare" in at-fault headlines can emphasize the cyclists' perceived mistakes or irresponsibility. Such linguistic choices not only amplify the severity of incidents but also frame the cyclist in a vulnerable position, potentially magnifying their fault or recklessness. This highlights the impact of emotive language on public engagement, with headlines potentially altering public reactions and creating echo chambers. These findings emphasize the crucial role of media language in framing cyclists' actions and influencing societal views on cycling safety~\cite{tan2016lost}.

\begin{figure}[t]
    \centering
    \includegraphics[width=\linewidth]{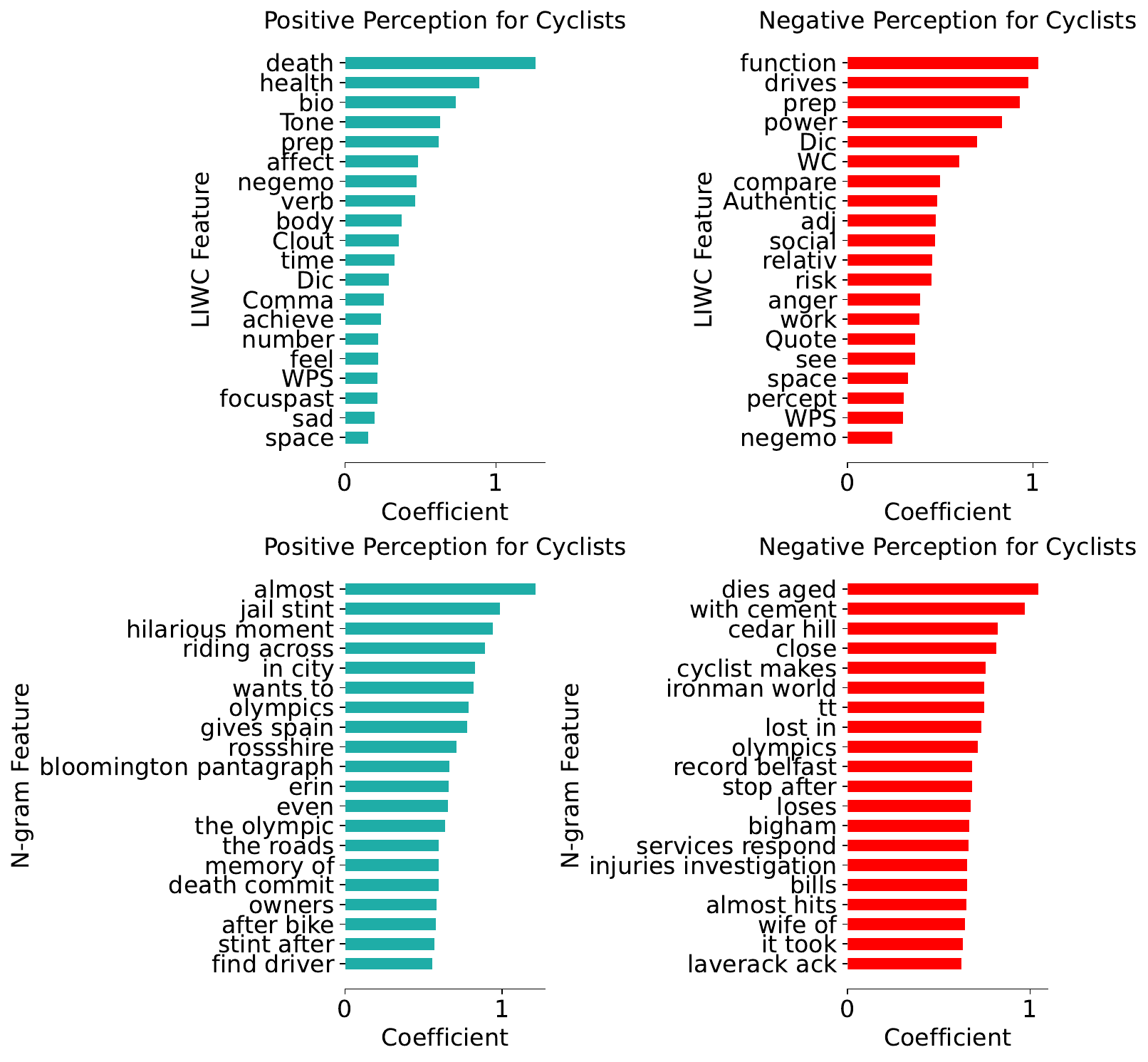}
    \caption{Top 20 Linguistic Factors (LIWC and n-grams (n=1,2)) in news headlines for positive or negative perception towards cyclists.\vspace{-1em}}
    \label{fig:lex-n-gram-perception}
\end{figure}


\vspace{1mm} \noindent \textbf{Linguistic Content for Cyclists At-Fault.} In the analysis of linguistic content for at-fault cyclists, n-grams like ``record belfast,'' ``risks friday,'' and ``cyclist rushed'' indicate cyclists' involvement in accidents, potentially as primary causes. Context-specific n-grams such as ``adidas goes'' and ``olympic medal'' relate to competitive events like the Olympics, framing cyclists in a sports context. N-grams like ``bike cyclistby,'' ``clipperd,'' and ``park austin'' reveal demographic and geographic aspects, indicating that news on cycling accidents often focuses on specific age groups or locations, potentially leading to stereotyping of these groups.

\vspace{1mm} \noindent \textbf{Linguistic Style for Negative Perception of Cyclists.} As shown in Figure~\ref{fig:lex-n-gram-perception}, the use of LIWC features like ``function,'' ``drives,'' ``prep,'' ``power,'' and ``risk'' in headlines about cycling incidents underscores potential dangers and casts cycling as perilous. Terms like "compare" and "anger" further amplify this negative framing. For example, the inclusion of words associated with negative emotions and critical assessments can lead to a perception of cycling as risky and cyclists as reckless.  

\vspace{1mm} \noindent \textbf{Linguistic Content for for Negative Perception of Cyclists.} Additionally, the use of specific context-related n-grams, such as ``dies aged,'' ``cyclist makes,'' and ``ironman world,'' highlights the severity and dramatic nature of these incidents, which can further skew public perception negatively. These linguistic features collectively convey negativity and biases, significantly impacting public perception and discourse around cycling incidents.

\section{LIMITATIONS AND FUTURE WORK}

Our model, primarily analyzing news headlines, faces limitations like not providing full story context and potentially featuring "click-bait" headlines, especially on social media. Extending the analysis to short articles or tweets could offer a more comprehensive view of public perceptions. Its English-centric and US-annotator-based approach may not accurately capture global perceptions, necessitating careful application in diverse contexts. Our research acknowledges the risk of news headline framing, particularly on sensitive topics, being misused to negatively influence public perception, highlighting the need for responsible use of our analysis tools to prevent bias and stereotypes. Future studies will expand beyond our current Google News API dataset to more comprehensively analyze temporal changes, regional differences, and political influences in cyclist news framing, aiming for a broader community impact.

\section{CONCLUSION}
In this paper, we make several contributions. First, we developed a new dataset for urban informatics to analyze how news headlines portray cyclists. Second, our novel BikeFrame Chain-of-Code approach, which jointly predicts multitask outcomes and incorporates a news source analysis module, significantly improves model performance. Third, we conducted an in-depth analysis of how cyclists are portrayed in the news. Key findings include general news websites reporting more accidents than cycling-specific sites, despite the higher fatality rates associated with cycling. Additionally, our study on gender-specific pronouns in news revealed only a slight difference in accident reporting between male and female mentions(20\% relative difference), contrasting with the actual higher accident rates for men (i.e., $>$700\% relative difference), which can exaggerate the dangers of cycling for women and potentially limit the number of female cyclists.

\bibliography{aaai22}

\appendix

\section{Paper Checklist to be included in your paper}

\begin{enumerate}

\item For most authors...
\begin{enumerate}
    \item  Would answering this research question advance science without violating social contracts, such as violating privacy norms, perpetuating unfair profiling, exacerbating the socio-economic divide, or implying disrespect to societies or cultures?
    Yes. However, there could be tools developed that generate headlines that are anti-cycling on purpose, particularly for groups of people using the dataset we developed.
  \item Do your main claims in the abstract and introduction accurately reflect the paper's contributions and scope?
    Yes
   \item Do you clarify how the proposed methodological approach is appropriate for the claims made? 
    Yes. We develop a novel multi-task learning method to take advantage of class relationships.
   \item Do you clarify what are possible artifacts in the data used, given population-specific distributions?
    Yes. But, not all population information is available. But, we do provide what we can regarding whether the content is US-based.
  \item Did you describe the limitations of your work?
    Yes
  \item Did you discuss any potential negative societal impacts of your work?
    Yes, see the ``LIMITATIONS AND FUTURE WORK'' section.
      \item Did you discuss any potential misuse of your work?
    Yes, see the ``LIMITATIONS AND FUTURE WORK'' section.
    \item Did you describe steps taken to prevent or mitigate potential negative outcomes of the research, such as data and model documentation, data anonymization, responsible release, access control, and the reproducibility of findings?
    Yes for what is related to this project, see the data description and annotation guideline sections.
  \item Have you read the ethics review guidelines and ensured that your paper conforms to them?
    Yes
\end{enumerate}

\item Additionally, if your study involves hypotheses testing...
\begin{enumerate}
  \item Did you clearly state the assumptions underlying all theoretical results?
    NA
  \item Have you provided justifications for all theoretical results?
     NA
  \item Did you discuss competing hypotheses or theories that might challenge or complement your theoretical results?
     NA. We have research questions to support our modeling and dataset results. But, these are not based on social theories in traditional hypothesis research.
  \item Have you considered alternative mechanisms or explanations that might account for the same outcomes observed in your study?
    NA
  \item Did you address potential biases or limitations in your theoretical framework?
    NA
  \item Have you related your theoretical results to the existing literature in social science?
    NA
  \item Did you discuss the implications of your theoretical results for policy, practice, or further research in the social science domain?
    NA
\end{enumerate}

\item Additionally, if you are including theoretical proofs...
\begin{enumerate}
  \item Did you state the full set of assumptions of all theoretical results?
   NA
	\item Did you include complete proofs of all theoretical results?
  NA
\end{enumerate}

\item Additionally, if you ran machine learning experiments...
\begin{enumerate}
  \item Did you include the code, data, and instructions needed to reproduce the main experimental results (either in the supplemental material or as a URL)?
   Yes. We included the main model and data in the supplementary material. The rest of the code will be released via GitHub upon acceptance.
  \item Did you specify all the training details (e.g., data splits, hyperparameters, how they were chosen)?
     Yes. See the METHODOLOGY section.
     \item Did you report error bars (e.g., with respect to the random seed after running experiments multiple times)?
    No.
	\item Did you include the total amount of compute and the type of resources used (e.g., type of GPUs, internal cluster, or cloud provider)?
    Yes. See the METHODOLOGY section.
     \item Do you justify how the proposed evaluation is sufficient and appropriate to the claims made? 
    Yes
     \item Do you discuss what is ``the cost`` of misclassification and fault (in)tolerance?
   Yes.
  
\end{enumerate}

\item Additionally, if you are using existing assets (e.g., code, data, models) or curating/releasing new assets, \textbf{without compromising anonymity}...
\begin{enumerate}
  \item If your work uses existing assets, did you cite the creators?
    Yes (pre-trained models)
  \item Did you mention the license of the assets?
    Yes, it is released under the Creative Commons Attribution License (CC BY 4.0).
  \item Did you include any new assets in the supplemental material or as a URL?
    Yes. The data is in the supplementary material.
  \item Did you discuss whether and how consent was obtained from people whose data you're using/curating?
   No.
  \item Did you discuss whether the data you are using/curating contains personally identifiable information or offensive content?
    No.
\item If you are curating or releasing new datasets, did you discuss how you intend to make your datasets FAIR (see \citet{fair})?
Yes. See the DATA COLLECTION AND ANNOTATION section.
\item If you are curating or releasing new datasets, did you create a Datasheet for the Dataset? 
No.
\end{enumerate}

\item Additionally, if you used crowdsourcing or conducted research with human subjects, \textbf{without compromising anonymity}...
\begin{enumerate}
  \item Did you include the full text of instructions given to participants and screenshots?
    Yes.
  \item Did you describe any potential participant risks, with mentions of Institutional Review Board (IRB) approvals?
     No, because our study was conducted internally using research assistants and was deemed not to require IRB approval.
  \item Did you include the estimated hourly wage paid to participants and the total amount spent on participant compensation?
    All graduate students are paid as research assistants which includes yearly stipend and tuition.
   \item Did you discuss how data is stored, shared, and deidentified?
   NA
\end{enumerate}

\end{enumerate}

\section{Appendix}
\section{A.1. Baseline Model Descriptions}

Likewise, we experiment with two novel multi-task learning methods to incorporate co-occurrence information into the model: Multi-Task RoBERTa (MTR) and Multi-Task RoBERTa using the Labeled Powerset Transformation (MTRLP). 

\vspace{1mm} \noindent \textbf{Random Baselines.} We use two random baselines from the scikit-learn package~\cite{pedregosa2011scikit}: Uniform and Stratified. The Uniform baseline makes predictions for each class with equal proportions. The stratified random baseline makes predictions based on the class proportions in the training dataset.

\vspace{1mm} \noindent \textbf{Linear SVM and Logistic Regression.} We trained a Linear SVM and LR model using frequency-inverse document frequency-weighting (TF-IDF)~\cite{schutze2008introduction} of unigrams and bigrams, a technique that assigns importance to words in a text corpus. For Linear SVM, we optimized the model by experimenting with different ``C'' values and L2 regularization. For LR, we used L1 regularization, `liblinear' solver, and 'balanced' class weight. Both models were implemented using scikit-learn library~\cite{pedregosa2011scikit}.

\vspace{1mm} \noindent \textbf{RoBERTa.}  We fine-tuned the RoBERTa~\cite{liu2019roberta} model from Huggingface~\cite{wolf2019huggingface}, averaging the second-to-last layer's token embeddings and passing them to a softmax layer for up to 25 epochs. The best model was selected using validation data. We employed cross-entropy loss, a mini-batch size of 8, a learning rate of 2e-5, and the Adam optimizer with a Cosine linear learning rate scheduler, without warm-up steps.


\vspace{1mm} \noindent \textbf{Multi-Task RoBERTa (MT RoBERTa).}  We employ a standard method of multi-task learning approach that is successful in similar text classification tasks~\cite{kochkina2018all}. Intuitively, the classes are related to each other. Hence, if we explore methods to take advantage of the implicit relationships between the classes.  Specifically, let $\mathbf{h} \in \mathbb{R}^k$ be the representation returned using RoBERTa where $k$ is the size of the hidden layer. Given $\mathbf{h}$, we jointly train the two output layers for Perception and Accident information (Related to an Accident and Fault). The RoBERTa parameters are shared for each task. To train the model, we simply perform a weighted average of the Cross-Entropy losses defined as
\begin{equation*}
    L = \sum_{i=1}^T w_i CE(\mathbf{y}_i, \mathbf{\hat y}_i)
\end{equation*}
where $T$ is the number of Tasks, $\mathbf{y}_i$ is the ground-truth classes for task $i$, $\mathbf{\hat y}_i$ are the predictions, $CE()$ is the Cross Entropy loss, and $w_i$ is a weight for task $i$. After experimenting with various weights for $w_i \in [0,1]$ we found weights of 1 to perform the best on our validation data.

\vspace{1mm} \noindent \textbf{Multi-Task RoBERTa using the Labeled Powerset Transformation (MTLPT RoBERTa).} While MT RoBERTa implicitly incorporates label co-occurrence information by jointly training each class, it does not explicitly capture the information which can limit co-occurrence knowledge acquisition for infrequently co-occurring classes. For multi-label classification, methods have been proposed to take advantage of class co-occurrence~\cite{tsoumakas2009mining}. The Labeled Powerset (LP) Transformation is a common approach~\cite{read2008pruned}. 

Intuitively, instead of training an output layer that predicts each class independently, e.g., ``Positive Perception'' and ``Cyclist-Fault'', a new class is created that combines all of the classes assigned to each instance. In this example, if the Perception is Positive and the Fault is assigned to the Cyclist, the new class would be ``Positive-Perception\_Cyclist-Fault'', which consists of transforming a multi-label problem into a single-label multi-class problem. In the transformed problem, each combination of labels presented in the original dataset is transformed into a single class. Despite the disadvantage of being the worst-case computational complexity (involving $2^L$ classes in the transformed multi-class problem where $L$ is the total classes), the LP transformation is simple, considers label correlations, and after transformation, any multi-class algorithm can be used for classification. However, many of the newly generated classes appear to infrequently make adequate predictions. Hence, we use this task as an additional multi-task regularizer. Specifically, we use the LP transformation as an auxiliary output layer trained in a multi-task setting. The auxiliary output layer is not used for inference. 

Formally, we define the new multi-task classification task as
\begin{equation*}
    L_i = CE(\mathbf{y}_i, \mathbf{y}_i) + \alpha CE(\mathbf{y}_p, \mathbf{\hat y}_p)
\end{equation*}
where $\alpha$ is a hyperparameter for the weight of the LP transformed loss, $\mathbf{y}_p$ is the vector of ground-truth for the LP transformed classes, and  $\mathbf{\hat y}_p$ represents the predictions. We train a model for each class independently (i.e., $L_i$ is a loss function for task $i$) where the LP transformation outputs are used as a regularizer. We also experiment with combining the LP transformation loss with the MT RoBERTa model described above. Overall, the final loss is defined as
\begin{equation*}
    L = \alpha CE(\mathbf{y}_p, \mathbf{\hat y}_p) + \sum_{i=1}^T w_i CE(\mathbf{y}_i, \mathbf{\hat y}_i) 
\end{equation*}
where, again, $\alpha$ is a hyperparameter and $w_i$ is the weight of each of the individual task losses. Empirically, we found setting everything to one (i.e., $\alpha$ and $w_i$) resulted in the best performance on the validation dataset.


\section{A.2. Summary Statistics for Male vs Female Articles for Unlabeled Corpus}
\begin{table}[h!]
\centering
\begin{tabular}{lcc}
\toprule
\textbf{Category} & \textbf{Male} & \textbf{Female} \\
\midrule
\textbf{Total Articles} & 966 & 603 \\
\textbf{Yes} & 588 & 288 \\
\textbf{No} & 378 & 315 \\
\textbf{Fault Attribution} & & \\
- Cyclist & 220 & 112 \\
- Other & 359 & 172 \\
- Unknown & 387 & 319 \\
\textbf{Perception} & & \\
- Negative & 405 & 217 \\
- Neutral & 359 & 248 \\
- Positive & 202 & 138 \\
\bottomrule
\end{tabular}
\caption{Summary Statistics for Male vs Female Articles, out of a total of 11,385 headlines from US-based websites}
\label{table:summary}
\end{table}

\section{A.3. Prompts Used in Our Experiments}

\vspace{2mm} \noindent \textbf{Prompt for Execute BikeFrame Chain-of-Code.\footnote{The DESIGN prompt is included as supplementary material with the dataset because of size.}}

\begin{tcolorbox}[colback=white, colframe=black, title=Bike Frame Execute]
```python\\
input\_text = ``{title}''\\
publisher\_title =  ``{publisher\_title}''\\
bike\_frame = BikeFrame(input\_text, publisher\_title)\\ \\
final\_answer = bike\_frame.analyze\_headline()\\ \\
print(``Final answer:''+ final\_answer)
``` 
\\ \\
\# Instruction: 
Generate the expected execution output (output from all print() functions) of the code. 
You don't have to actually run the code and do not care about 'not implemented error'.
\end{tcolorbox}

\vspace{2mm} \noindent \textbf{Prompt for Direct Prompting}
\begin{tcolorbox}[colback=white, colframe=black, title=Direct Prompting]
\#\#\# System: Reads a given input text and analyzes news headlines about cycling by determining whether an accident is mentioned, identifying the party at fault, and assessing the readers' perception towards cyclists. The final answer should be formatted as follows: "Final answer: Accident: [Yes/No], Fault: [Cyclist/Other/Unknown], Perception: [Negative/Positive/Neutral]."

\#\#\# User: text for the task: \{title\} 
Final answer should be at the end of your answer and its format should be like ``Final answer: your\_answer''. 
Generate output following the task description above. 
Output: 
\end{tcolorbox}

\vspace{2mm} \noindent \textbf{Prompt for Zero-shot CoT}
\begin{tcolorbox}[colback=white, colframe=black, title=Zero-shot CoT]
\#\#\# System: Reads a given input text and analyzes news headlines about cycling by determining whether an accident is mentioned, identifying the party at fault, and assessing the readers' perception towards cyclists. The final answer should be formatted as follows: "Final answer: Accident: [Yes/No], Fault: [Cyclist/Other/Unknown], Perception: [Negative/Positive/Neutral]."

\#\#\# User: text for the task: \{title\} 
Final answer should be at the end of your answer and its format should be like ``Final answer: your\_answer''. 
Generate output following the task description above. 
Output:
Let’s think step by step.
\end{tcolorbox}

\end{document}